\documentclass[twocolumn, prl]{revtex4}

\usepackage{graphicx}
\usepackage{dcolumn}
\usepackage{bm}

\begin{document}

\title{ A Renormalization Group Method for Quasi One-dimensional
       Quantum Hamiltonians}

\author {S. Moukouri}

\affiliation{ Michigan Center for Theoretical Physics and
             Department of Physics, \\
         University of Michigan 2477 Randall Laboratory, Ann Arbor MI 48109}

\author{L. G. Caron}

\affiliation{ D\'epartement de physique, Universit\'e de Sherbrooke \\
    2500 Bd. de l'Universit\'e, Sherbrooke, J1K 2R1, Qu\'ebec, Canada}

\begin{abstract}
 A  density-matrix renormalization group (DMRG) method for highly anisotropic
two-dimensional systems is presented. The method consists in
applying the usual DMRG in two steps. In the first step, a pure
one dimensional calculation along the longitudinal direction is
made in order to generate a low energy Hamiltonian. In the second
step, the anisotropic 2D lattice is obtained by coupling in the
transverse direction the 1D Hamiltonians. The method is applied to
the anisotropic quantum spin half Heisenberg model on a square
lattice.
\end{abstract}

\maketitle

The density-matrix renormalization group method introduced by
White \cite{white} a decade ago has proven a remarkable efficiency
in the computation of the ground-state
\cite{white2,moukouri_KONDO, DMRG_book}, thermodynamic
\cite{moukouri_THERMO,bursill} and dynamic \cite{hallberg,shibata}
properties of one-dimensional interacting electron models as well
as electron-phonon systems \cite{caron,zhang}. An extension of the
method to two dimensional systems is currently the object of some
efforts \cite{xiang}. But an efficient 2D algorithm is still
lacking.

An interesting situation which has been so far overlooked  is the possible
application of the DMRG technique to strongly anisotropic 2D systems.
In such systems,  the interaction Hamiltonian can be written as the sum of two
terms: $H=H_{\parallel}+gH_{\perp}$; where $H_{\parallel}$ is a the sum over
one-dimensional (1D) Hamiltonians (longitudinal direction) and $H_{\perp}$ is
the interaction between these 1D systems (transverse direction).
$H_{\parallel}$ and $H_{\perp}$ are of the same magnitude and $g \ll 1$.
This type of situation arises in various physical systems including quasi 1D
organic and inorganic materials, weakly coupled ladder systems, carbon
nanotube ropes, quasi 1D antiferromagnets, quasi 1D spin-Peierls systems, etc.
It is the purpose of this letter to show that a DMRG algorithm can be
implemented for the study of such anisotropic models.

The standard approach to anisotropic Hamiltonians such as $H$ is
to first diagonalize $H_{\parallel}$ and then use its eigenstates
to compute the corrections introduced by $H_{\perp}$. We present
below an algorithm  which follows this spirit. This method
consists in applying a two-step DMRG method to Hamiltonians having
the form of $H$. The standard DMRG is first used along the
longitudinal direction until the scale of $gH_{\perp}$ is reached.
Then a transformation is made in order to keep only a small number
( a few tens) of low lying states of $H_{||}$. These states are
then used as a starting point for a second DMRG calculation along
the transverse direction. The validity of this procedure is
discussed. As in  the thermodynamic algorithm
\cite{moukouri_THERMO} which was introduced by the authors, the
accuracy depends on not only the number of states $m$ kept by
blocks, but also on the number of the target states $M$.

\begin{figure}
\includegraphics[width=2 in]{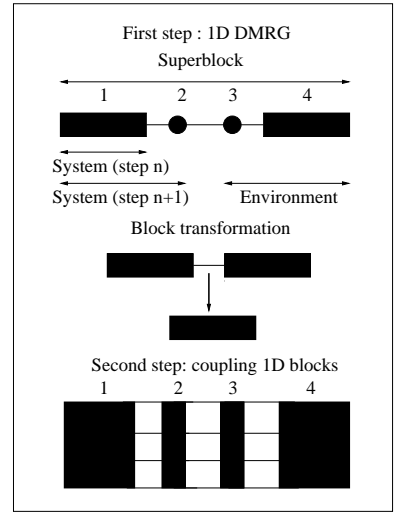}
\caption{Sketch of the two-step density-matrix renormalization group algorithm}
\label{superblock}
\end{figure}

The two-step DMRG is illustrated in Fig.~\ref{superblock}. The
first step of the quasi 1D algorithm is the usual 1D DMRG method.
For the sake of completeness, we briefly recall the main steps of
the algorithm. For more details, we refer the reader to the more
extensive presentations in ref\cite{DMRG_book}. An iteration of
the DMRG algorithm with the infinite system method proceeds as
follows: (a) the Hamiltonian for the superblock 1+2+3+4 ( where
the blocks 1 and 4 come from previous iterations and 2 and 3 are
new added ones.  Blocks 2 and 3 are usually single site blocks)
 is diagonalized in order to obtain the few $M$ lowest lying wave functions
$| \Psi_{k}(1,2,3,4) \rangle$, $k=1,M$. (b) the reduced density matrix of
blocks 1+2 is constructed. This reduced density matrix is
$\rho =\sum_{k}\omega _{k} \rho _{k}$, where $\rho_{k}$ and $\omega_{k}$ are
respectively the reduced density matrix of the $k^{th}$ state and its weight.
$\rho_{k}$ is related to the superblock wave function $\Psi_k(1,2,3,4)$
through the relation, $\rho_{k}(1,2;1',2') = \sum_{3,4} |\Psi_{k}(1,2,3,4)
\rangle \langle \Psi_{k}^{*}(1',2',3,4)|$. $\rho$ is diagonalized and the $m$
eigenstates with the highest eigenvalues $\lambda_{\alpha}$ are kept.
Since $\sum_{\alpha} \lambda_{\alpha}=1$, because $\rho$ is a density matrix,
the error made by truncating this sum to $m$ states is $p_m=1-\sum_{\alpha=1,m}
\lambda_{\alpha}$. (c) These states form a new reduced basis in which all the
operators have to be expanded and the block 1+2 is renamed as block 1. New
block 2 and 3 are added to form a new superblock. This procedure is repeated
until one reaches the desired lattice size. This means typically, once the
energy separation between the eigenstates of the 1D system are of the same
magnitude as g.

 Once the size $L+2$ of the superblock is large enough, {\it i.e} when the
difference between the first excited state and the ground state is such that
$|E_0 - E_1| \approx g$, we perform a block transformation, of the type of
the old RG method \cite{weinstein,jullien}, on blocks 1 and 4.
The two $L/2$-lattices with $m$ states are coupled, then renormalized to a
$L$-lattice with $m$ states. These $m$ states are the ground state and the
low-lying excited states of the $L$-lattice. They can now be coupled
 through $g$. One can note that the boundary condition problem in
 the block RG method raised by White and Noack \cite{white-noack} does
not exist here, this is  because the block transformation is only applied once.
A 2D lattice is then generated by using the 1D density-matrix renormalization
group algorithm  with chains as units instead of sites.

This new block 1 will serve, in the second DMRG step, as a
building unit for a new series of iteration that are similar to
those performed in step 1.  This block is labelled block 2 in the
second step since it plays the role played by a single site in
step 1. This two-step method can be justified as follows: in the
study of the ground state properties of a quasi 1D system, since
$g \ll 1$, $g H_{\perp}$ may be neglected in the early steps of
the renormalization process. The reason is that finite size
effects act as temperature. When the size is small, the excitation
energies are far greater than the transverse coupling, $g \ll |E_0
- E_n|$, therefore $g$ is irrelevant.

 We now illustrate the two-step DMRG method in the case of weakly coupled
spin-half Heisenberg chains on a square lattice. The Hamiltonian
reads,

\begin{eqnarray}
H_{spins}=\sum_{i,l}{\bf S}_{i,l}{\bf S}_{i+1,l}+J_{\perp}
\sum_{i,l}{\bf S}_{i,l}{\bf S}_{i,l+1}
\label{heisenberg}
\end{eqnarray}

\noindent where the ${\bf S}_{i,l}$ are the usual spin-half operators. 
Following the
prescriptions above, we start by applying the 1D DMRG on the single chain
Hamiltonian $H_{\parallel,l}= \sum_{i}{\bf S}_{i,l}{\bf S}_{i+1,l}$.
 It was shown by White \cite{white} that a relatively small number of $m$
states kept in block 1 can lead to an astonishingly high accuracy for the
ground-state energy and spin correlation functions.
An accuracy  ranging from $10^{-6}$ to better than $10^{-10}$ for the
ground state energy of a $L=28$ sites chain was obtained by keeping only
$m=16$ to $m=64$ states with modest computational power.

A set of these highly accurate low energy eigenstates $|\phi_{n_l}>$ and
eigenvalues $\epsilon_{n_l}$ of $H_{\parallel,l}$ is first obtained by the 1D
DMRG. The eigensets of the disconnected chain Hamiltonian
 $H_{\parallel}= \sum_l H_{\parallel,l}$ are given by the relations:

\begin{table}
\begin{ruledtabular}
\begin{tabular}{ccccc}
 $m$ & 1D & Ladder 1 &Ladder 2 & 2D \\
\hline
 16 &-13.98792  & -13.62493 & -14.00113 & -14.07066 \\
 24 & -13.99650 & -13.69023 & -14.01248 & -14.08803 \\
 32 & -13.99724 & -13.70638 & -14.01526 & -14.09810 \\
 40 & -13.99729 & -13.71446 & -14.01749 & -14.10489 \\
\end{tabular}
\end{ruledtabular}
\caption{ Ground state energies per chain of single chain (1D) with $L=32$,
of a two-leg ladder $2 \times L$ obtained with the usual DMRG (Ladder 1), of the
 same ladder obtained with the two-step method (Ladder 2), of a $32 \times 32$
 system (2D).}
\label{table1}
\end{table}

\begin{eqnarray}
\Phi_{[n]}= \prod_{l} \otimes |\phi_{n_l}>,\ E_{[n]}=\sum_{l} \epsilon_{n_l}
\end{eqnarray}

\noindent $[n]=(n_1,n_2,...,n_L)$, $n_{l}$ corresponds to an eigenset on
the chain $l$. The eigenset $ \Phi_{[n]}$, $E_{[n]}$ diagonalizes
$H_{\parallel}$:

\begin{eqnarray}
<\Phi_{[n]}|H_{\parallel}|\Phi_{[m]}>=E_{[n]} \delta_{[n],[m]}.
\label{h_0}
\end{eqnarray}

\noindent The matrix elements of $H_{\perp}$ in the basis formed by the
$\Phi_{[n]}$ are:

\begin{eqnarray}
<\Phi_{[n]}|H_{\perp}|\Phi_{[m]}>= \sum_{l} <\Phi_{[n]}|\sum_i {\bf S}_{i,l}
{\bf S}_{i,l+1}|\Phi_{[m]}>.
\label{h_perp}
\end{eqnarray}

This last equation ~\ref{h_perp} can readily be transformed to:
\begin{eqnarray}
<\Phi_{[n]}|H_{\perp}|\Phi_{[m]}> = \sum_{il}  {\bf \tilde{S}}_{i,l}^{n_l,m_l}
{\bf \tilde{S}}_{i,l+1}^{n_{l+1},m_{l+1}}.
\label{h_perp2}
\end{eqnarray}

\noindent where ${\bf \tilde{S}}_{i,l}^{n_l,m_l}=<\phi_{n_l}|{\bf S}_{i,l}|\phi_{m_l}>$.
The renormalized Hamiltonian $ \tilde {H}$ in the basis of the $\Phi_{[n]}$ is
thus:

\begin{eqnarray}
\tilde{H} \approx \sum_{[n]} E_{[n]} |\Phi_{[n]}><\Phi_{[n]}| +
 J_{\perp} \sum_{il} {\bf \tilde{S}}_{i,l} {\bf \tilde{S}}_{i,l+1}.
\end{eqnarray}

\begin{figure}
\includegraphics[width=3. in]{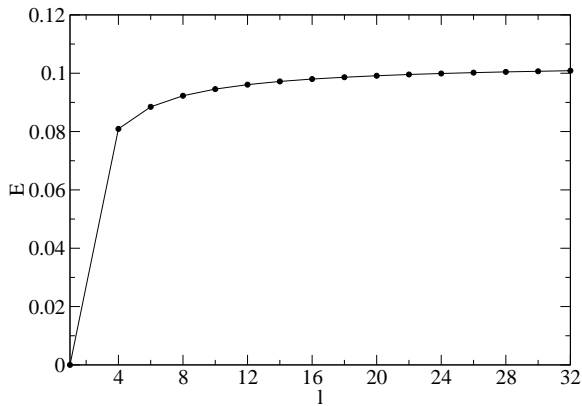}
\caption{The binding energy of the $32 \times l$ system (relative to
the isolated 32 site chain) for $m=32$ and $J_{\perp}=0.1$ }
\label{groundstate}
\end{figure}

$\tilde {H}$ is studied by the usual 1D DMRG algorithm. It is however to be
noted that the computational requirement for $\tilde {H}$ are greater than
in the study of a simple spin chain performed in the first step. The
renormalized spin operators ${\bf \tilde{S}}_{i,l}$ are now $m \times m$
matrices instead of $2 \times 2$. The size of the superblock in the second
step is $m'^2 \times m^2$ (it is $4 \times m^2$ in step 1) if $m'$ states
are kept in the two external blocks. We were nonetheless able to
perform the calculations on a workstation for up to $m = 48$.

\begin{table}
\begin{ruledtabular}
\begin{tabular}{cccc}
 $m$ & $\Delta(L)$ & $\delta E(L)$ &$dim S_z=0$  \\
\hline
 16 & 0.11671 & 0.58664 & 12 251   \\
 24 & 0.11780 & 0.65044 & 66 850   \\
 32 & 0.11777 & 0.72977 & 205 920  \\
 40 & 0.11774 & 0.78363 & 477 670   \\
\end{tabular}
\end{ruledtabular}
\caption{ Finite size spin gap $\Delta(L)$, band-width of the retained states
 $\delta E(L)$ and the dimension of the subspace with $S_z=0$ in the two-step
 method.}
\label{table2}
\end{table}

It is instructive to compare the results which are obtained by the
proposed method with a known case. A simple situation is the
two-leg spin ladder model. We have computed the ground state
energy of the ladder system in two ways. First, we use the usual
DMRG algorithm which have been widely applied for the study of
this system during the last decade \cite{azzouz}. Second, we
compute the ground-state energy of the two leg ladder by first
obtaining the low energy Hamiltonian of a single chain. For this,
we targeted $M=5$ states corresponding to the lowest states of the
spin sectors $S_z= \pm 2$, $ \pm 1$ and $0$. It is necessary to
target other spin sectors when building the low energy Hamiltonian
because, when forming the tensor product, they may lead to lower
energy states than some $S_z=0$ states. A criteria for the
selection of spin sector to be selected is similar to the one
applied in the thermodynamic algorithm. If the lowest state of
$S_z$ is higher than the highest state retained in $S_z=0$, the
subspace $S_z$ is not retained. We then coupled two low energy
1D Hamiltonians with $J_{\perp}$ to form a ladder. This
corresponds to performing the second step of the two-step DMRG
with one iteration and two blocks instead of four.

In table~\ref{table1} we compare the ground state energies of the
ladder obtained by the two methods  as well as those of a single
chain for $L=32$. The infinite system DMRG method with open boundary 
conditions was used. The energy per site of the $32$ site single chain 
for $m=40$ is $-0.4374$ which is still far from the exact ground state 
energy of the infinite chain $-0.4431$. This is because chains with open 
boundary conditions converge very slowly (as $1/L$) to the thermodynamic 
limit. It can be seen that the energy of the two step DMRG are systematically
better than those of the simple DMRG. The latter are found to be
even higher than those of a single chain. This can be understood
by the fact that the DMRG method is less accurate in situations
where the ground state is nearly degenerate. This type of
situation is, for instance, encountered in the small coupling
region of the Kondo lattice \cite{ moukouri_KONDO} where a large
number of states must be kept to get reliable results. Another
look at this is that when $J_{\perp}$ is small, the two chains are
nearly independent, the $m$ states are shared between the two
chains $\sqrt{m}$ states for each chain \cite{caron2}.

\begin{figure}
\includegraphics[width=3. in]{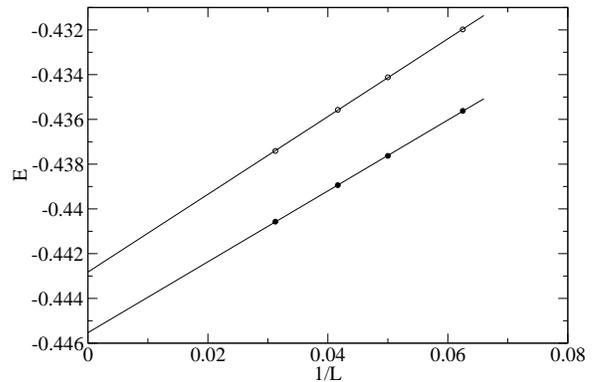}
\caption{The ground state energy per site for the 2D lattice for $J_{\perp}=0$
(open circles) and $J_{\perp}=0.1$ (filled circles) as a function of $L$ for
 $m=32$.}
\label{gsfit}
\end{figure}


 The ground state energy of the two dimensional system ($32 \times 32$) is
shown in the last column of table~\ref{table1} and Fig.~\ref{groundstate} shows
the variation of the ground state energy per chain as a function of the
chain number $l$, the reference is a single chain with $L=32$. It displays a
rapid convergence as $l$ increases. We kept the same number of
states in the two external blocks as in the internal blocks during the second
step of the DMRG. Truncation errors are between $10^{-3}$ and $10^{-5}$ for
 $m=16$ to $40$. One should note, however, that these errors are relative to
the truncated Hamiltonian. Hence they do not necessarily represent
the accuracy of the method. It is important to verify that the
truncated Hamiltonian generated for the single chain is accurate
enough, for the ground state and for the low lying state, to be
used as a building block for the 2D lattice. This will depend
indeed on $m$ and $M$, but also on $J_{\perp}$ and $L$. One can
easily see that for a fixed $L$ and $J_{\perp} \ll
\Delta_{\sigma}(L)$, $\Delta_{\sigma}(L)$ is the finite size spin gap, 
the interchain matrix elements will be
negligible. The system will behave as a collection of free chains
even if $J_{\perp}$ is turned on. Now if $J_{\perp} \sim \delta
E(L)$, $\delta E(L)$ is the width of the retained states, 
the matrix elements of the states having higher energy
which have been truncated out have a non-negligible contribution.
We thus expect the two-step DMRG technique to be reliable when
$J_{\perp}$ is within the range $\Delta_{\sigma}(L) \simeq
J_{\perp} \ll \delta E(L)$. Values of  $\Delta_{\sigma}(L)$ and  $\delta E(L)$
 are displayed in table~\ref{table2}.  An estimate of the extrapolated ground
state energy per site for the 2D system (Fig.~\ref{gsfit}) is -.4455 for 
$J_{\perp}=0.1$ when $m=32$ (the same value of $m$ gives -.4428 for 
$J_{\perp}=0$). This is to be compared to the quantum Monte Carlo (QMC) 
estimate -.4485 for the infinite system obtained using clusters with periodic 
boundary conditions \cite{alvarez}.

\begin{figure}
\includegraphics[width=3. in]{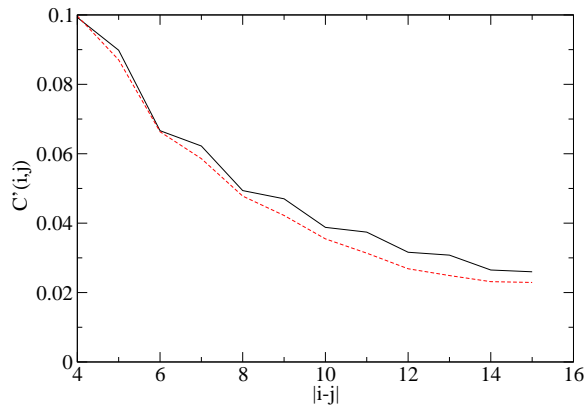}
\caption{The averaged ground state correlation function
  $C'(i,j)=(-1)^{|i-j|}<{\bf S}_i {\bf S}_j>$ of a single chain
 (full) and of the $32 \times 32$ system (dashed) for $m=32$ and $J_{\perp}=0.1$ }
\label{avspincor}
\end{figure}

The possible occurrence of long-range order is studied by
observing the behavior of spin correlation functions. This is
however complicated by the use of open boundary conditions. It is
known that \cite{white,caron2} the OBC induce a strong odd-even
alternation in the spin correlation functions. 
This alternation is also observed for the 2D system with
$J_{\perp}=0.1$. In order to avoid this alternation, we defined
the spin correlation function as the average between these two
states. This is done by computing $<{\bf S}_1 {\bf S}_l>$ in two
ways. ${\bf S}_1$ is taken as the origin of a strong link or as
the origin on a weak link. This averaging process reduces the
even-odd alternation as shown in Fig.~\ref{avspincor} for the
single chain; the alternation is suppressed in the 2D system. The
suppression of the alternation in the 2D system after averaging is
probably due to the fact that the valence bond picture is not
valid for the 2D system, even with open boundary conditions. The
observation of the alternation before averaging is simply
reminiscent of the starting point which is a single chain.

We have shown that the DMRG can successfully be applied to highly
anisotropic 2D systems. We used the spin-half Heisenberg
anisotropic model on a square lattice as a test case. Even in this
simple case, the computational power required increases
significantly with the number of block states kept as shown in the
last column of Table.~\ref{table2}. An analysis of the program
reveals that $75 $ percent of the CPU time is spent during the
calculation of the superblock wave function. This step involves
mostly matrix-vector products. A significant improvement in speed
is thus expected by the use of parallel computers. Furthermore,
the initial guess for the superblock wave function was a random
vector at each iteration. Another improvement of the program is to
use a transformed state from the previous iteration as done in the
standard DMRG method \cite{white-DAVIDSON}.

\begin{acknowledgements}
Part of this work was performed when S. M. was at the Groupe de
Physique des Solides, Universit\'e Paris VI. We wish to thank
Professor Claudine Nogu\'era and her group for their hospitality.
We thank J.V Alvarez and A. Sandvik for sharing their QMC data.
\end{acknowledgements}

\end{document}